\documentclass[prd,nofootinbib,preprintnumbers,twocolumn,showpacs,floatfix,superscriptaddress]{revtex4-1}

\usepackage[T1]{fontenc}
\usepackage[utf8]{inputenc}
\usepackage{lmodern, amsmath, graphicx, pifont, adjustbox, bm, xcolor}

\makeatletter\g@addto@macro\bfseries{\boldmath}\makeatother

\usepackage[colorlinks=true, citecolor=blue, urlcolor=blue, linkcolor=blue, breaklinks=true, pdfpagelabels=false]{hyperref}

\newcommand{\appendixref}[1]{\hyperref[#1]{appendix~\ref{#1}}}
\def\equationautorefname~#1\null{eq.\,(#1)\null}

\newcommand*{\sqb}[1]{[#1]}
\newcommand*{\anb}[1]{\langle#1\rangle}
\newcommand*{\la}{\langle}
\newcommand*{\ra}{\rangle}
\newcommand*{\lb}{\lbrack}
\newcommand*{\rb}{\rbrack}
\newcommand*{\bb}[1]{[ #1 ]}
\newcommand*{\bd}[1]{\langle  #1\rangle}

\newcommand*{\MM}{\mathcal{M}}

\newcommand*{\dop}{\ensuremath{\dim\{\text{operator}\}}}
\newcommand*{\dspin}{\ensuremath{\dim\{\text{spinors}\}}}
\newcommand*{\nt}{\ensuremath{n_\text{tensors}}}
\newcommand*{\nl}{\ensuremath{n}}
\newcommand*{\htot}{\ensuremath{h_\text{tot}}}
\newcommand*{\sym}[1]{\underline{#1}}

\begin{document}

\title{Enumerating higher-dimensional operators with on-shell amplitudes}

\author{Gauthier Durieux}
\affiliation{Physics Department, Technion -- Israel Institute of Technology, Haifa 3200003, Israel} 
\author{Camila S.\ Machado}
\affiliation{PRISMA$^+$  Cluster of Excellence \& Mainz Institute for Theoretical Physics, Johannes Gutenberg-Universit\"at Mainz, 55099 Mainz, Germany} 
\preprint{MITP/19-090}

\begin{abstract}
We establish a simple formula for the minimal dimension of operators leading to any helicity amplitude.
It eases the systematic enumeration of independent operators from the construction of massless non-factorizable on-shell amplitudes.
Little-group constraints can then be solved algorithmically for each helicity configuration to extract a complete set of spinor structures with lowest dimension.
Occasionally, further reduction using momentum conservation, on-shell conditions and Schouten identities is required.
A systematic procedure to account for the latter is presented.
Dressing spinor structures with dot products of momenta finally yields the independent Lorentz structures for each helicity amplitude.
We apply these procedures to amplitudes involving particles of spins $0,1/2,1,2$.
Spin statistics and elementary selection rules due to gauge symmetry lead to an enumeration of operators involving gravitons and standard-model particles, in the effective field theory denoted GRSMEFT.
We also list the independent spinor structures generated by operators involving standard-model particles only.
In both cases, we cover operators of dimension up to eight.
\end{abstract}

\maketitle

\section{Introduction}
Even before being known as such, effective field theories (EFTs) were employed in the early 1930s to model light-by-light scattering and nuclear beta decay.
Today, among a wide range of applications, EFTs are notably exploited to parametrize new physics hiding beyond the scales directly probed experimentally.
The lack of clear indications about the nature of physics beyond the standard model (SM) motivates the use of such model-independent approaches and the development of efficient methods to study the standard-model effective field theory (SMEFT).

Extremely successful in renormalizable theories, on-shell amplitude methods have also been applied to EFTs. 
Theories with enhanced soft limits have been the subject of various studies~\cite{Cheung:2016drk, Cheung:2018oki, Elvang:2018dco, Low:2019ynd}.
Non-renormalizations~\cite{Cheung:2015aba,Bern:2019wie} and non-interferences~\cite{Azatov:2016sqh} that are obscured by the Lagrangian formalism were exposed in the SMEFT.
The construction of a complete set of independent operators (i.e.\ a basis) can also be rather cumbersome in the usual Lagrangian approach.
Besides Hilbert series~\cite{Henning:2015alf, Henning:2015daa} and harmonics techniques applying to distinguishable particles~\cite{Henning:2019enq}, on-shell methods have also been employed in this task.
They have the advantage of avoiding the operator redundancies arising from field redefinitions.
The counting of operators relevant to amplitudes involving one colour singlet and up to three gluons was presented in ref.~\cite{Shadmi:2018xan}.
The enumeration of amplitudes corresponding to the full set of dimension-six SMEFT operators was carried out in ref.~\cite{Ma:2019gtx}.
The case of massive SM amplitudes was also addressed both at the renormalizable~\cite{Christensen:2018zcq, Christensen:2019mch, Bachu:2019ppp} and non-renormalizable~\cite{Aoude:2019tzn, Durieux:2019eor} levels.
All massive three-point and one four-point electroweak amplitudes were bootstrapped and their high-energy limits studied~\cite{Durieux:2019eor}; factorizable bosonic~\cite{Aoude:2019tzn, Bachu:2019ppp} and fermionic~\cite{Durieux:2019eor} four-point amplitudes were constructed; constraints due to electroweak symmetry breaking were observed to emerge from perturbative unitarity~\cite{Aoude:2019tzn, Durieux:2019eor, Bachu:2019ppp}; a matching to the SMEFT was carried out in the broken phase~\cite{Aoude:2019tzn, Durieux:2019eor}.

In this paper, we further advance the program of operator enumeration using massless on-shell amplitudes.
The case of interactions between gravitons and SM particles is examined as showcase example.
The Hilbert-series enumeration of operators of dimension eight at most~\cite{Ruhdorfer:2019qmk} is reproduced with on-shell methods.
Our technique also allows for a straightforward listing of independent spinor structures generated by SMEFT operators.
We limit ourselves to operators of dimension eight at most in this case too.

Massless on-shell amplitudes are reviewed in \autoref{sec:lg}.
An algorithmic procedure is notably described to construct spinor structures compatible with little-group constraints for a given helicity amplitude.
A systematic means of exhausting relations deriving from Schouten identities is also presented.
We establish a simple formula for the minimal dimension of operators leading to any helicity amplitude in \autoref{sec:min_op_dim}.
Applications to the GRSMEFT and SMEFT cases are treated in \autoref{sec:enumerations}.
Helicity amplitudes generated by operators of dimension eight at most are listed.
Minimal spinor structures compatible with little-group constraints are constructed systematically.
Their reduction to an independent set is carried out explicitly, before dressing them with additional dot products of momenta.
Constraints deriving from gauge invariance and spin statistics are accounted for in the GRSMEFT case.
Our results are summarized in \autoref{tab:amps} and \autoref{tab:sm_amps}.

\section{Massless amplitudes}
\label{sec:lg}\vspace*{-1mm}

In the on-shell formalism, massless amplitudes with $\nl$ external legs are constructed from (linear combinations of) square $[ij]$ and angle $\anb{ij}$ spinor brackets (see e.g.\ refs.~\cite{Mangano:1990by, Elvang:2015rqa, Schwartz:2013pla}) raised to integer powers that we denote $a_{ij}$ and $b_{ij}$, respectively:
\vspace*{-.5mm}
\begin{equation}
\mathcal{M}_\nl(h_1,...,h_n) \propto \prod_{i<j}  \;\:[ij]^{a_{ij}} \anb{ij}^{b_{ij}}\,.
\end{equation}
Spinor brackets have mass dimension $1$.
The $\nl(\nl-1)/2$ brackets of each type involve spinors of each possible pair of particles:
$
12,	13,	14,	\ldots,
	23,	24,	\ldots,
		34,	\ldots
$.
All momenta are conventionally taken to be either incoming or outgoing.
Little-group covariance requires that the helicity weights of the spinor structure match the helicity $h_i$ of the external particles.
This leads to $\nl$ constraints:
\begin{equation}
\sum_{i<j:\: i=k\text{ or }j=k}\hspace{-5mm} (a_{ij}-b_{ij}) = 2h_k\,,
\quad\text{for each } k=1,\ldots, n\,.
\label{eq:lg}
\end{equation}
Each $a_{ij},b_{ij}$ power appears in exactly two of those equations.
Summing them, one therefore obtains that:
\begin{equation}
\sum_{i<j} (a_{ij}-b_{ij}) = \sum_i h_i \equiv \htot,
\label{eq:tot_hel}
\end{equation}
where we have defined $\htot\equiv\sum_i h_i$, the total helicity.
On the other hand, the mass dimension of a spinor structure is given by the sum of all spinor bracket powers:
\begin{equation}
\dspin = \sum_{i<j}(a_{ij}+b_{ij}).
\label{eq:spin_dim}
\end{equation}
In massless three-point amplitudes (which can be written down for complex momenta), either the square or angle spinors are proportional to each other, leading to vanishing antisymmetric bracket contractions.
Massless three-point amplitudes can thus be constructed from a single type of brackets.
The constraints of \autoref{eq:lg} then form a linear system of three equations ($\nl=3$) with three unknowns ($\nl(\nl-1)/2=3$).
A well-known unique solution is thus found for either square or angle brackets:
\begin{align}
\label{eq:3pt-massless}
	&\MM_3(h_1,h_2,h_3)\propto \\
	&    \begin{cases}
	\,\bb{12}^{+h_1+h_2-h_3}\: \bb{23}^{-h_1+h_2+h_3}\: \bb{13}^{+h_1-h_2+h_3}\,,\\
	\bd{12}^{-h_1-h_2+h_3}\bd{23}^{+h_1-h_2-h_3} \bd{13}^{-h_1+h_2-h_3}\,.\nonumber
	\end{cases}
\end{align}
Given that a massless three-point amplitude only involves either square or angle brackets, \autoref{eq:tot_hel} and \autoref{eq:spin_dim} can be combined to obtain that $\dspin=\htot$ when square brackets are employed and  $\dspin=-\htot$ with angle brackets.
To preserve locality, a positive mass dimension is required.
One is therefore forced to respectively use square and angle brackets for positive $\htot>0$ and negative $\htot<0$ total helicities.

In this paper, we address the construction of non-factorizable amplitudes (having a trivial analytic structure with neither kinematic poles nor branch cuts) generated by contact operators.
Being local, they can always be written in a form that does not involve any negative power of spinor bilinears:\footnotemark
\begin{equation}
a_{ij},b_{ij}\ge 0\quad\forall ij,\quad\text{for non-factorizable amplitudes}.
\label{eq:non-fac}
\end{equation}
We moreover target the construction of spinor structures that cannot be reduced to simpler ones, involving fewer spinors.
\footnotetext[\thefootnote]{{Massless three-point amplitudes being unphysical, they need not be local but should only give rise to local four-point amplitudes once \emph{glued} together~\cite{McGady:2013sga}.
Apparent non-locality in massless three-point amplitudes is only allowed to appear in gauge couplings where it cancels in four-point amplitudes thanks to Lie algebra and Jacobi identities.
We omit them since our focus is on non-renormalizable operators.}}
Spinor structures involving positive powers of both $[ij]$ and $\anb{ij}$ are not, in this sense, minimal.
The $[ij]\anb{ji}=2\,p_i\cdot p_j\equiv s_{ij}$ equality indeed allows us to reduce the number of spinors they involve.
Solving little-group constraints for each $a_{ij}-b_{ij}$ difference, one would use a positive $a_{ij}$ power of the $ij$ square bracket for $a_{ij}-b_{ij}>0$ solutions (setting $b_{ij}=0$) and a positive $b_{ij}$ power of the $ij$ angle bracket for $a_{ij}-b_{ij}<0$ solutions (setting $a_{ij}=0$):
\begin{equation}
\begin{gathered}
a_{ij}=0\quad\text{or}\quad b_{ij} = 0\qquad\text{for each $ij$ pair,}
\\\text{in minimal spinor structures.}
\end{gathered}
\label{eq:minimal}
\end{equation}
For the minimal spinor structures of non-factorizable amplitudes involving $\nl$ legs, little-group constraints form a linear system of $\nl$ equations with $\nl(\nl-1)/2$ unknowns.
There are thus $\nl(\nl-3)/2$ parameters left undetermined: none for $\nl=3$, two for $\nl=4$, five for $\nl=5$, etc.
One can then for instance fix such a number of $a_{ij}-b_{ij}$ differences (spanning small integers is sufficient to obtain spinor structures of lowest dimension) and solve the linear system of little-group constraints for the others.

The spinor structures generated at this point are not necessarily independent.
Momentum conservation, on-shell conditions, and Schouten identities can be used to reduce them to an independent set in which no single spinor structure can be expressed as a linear combination of others.
In an amplitude with $\nl$ external legs, one Schouten identity can be written down for each possible set of four spinors (square or angle).
There are thus $\nl!/4!(\nl-4)!$ of them.
Each can be used to eliminate a fixed product of spinor bilinear.
In particular, in an amplitude featuring four legs (or four particles of non-vanishing helicity), one of  the three $\sqb{12}\sqb{34}, \sqb{13}\sqb{24}, \sqb{14}\sqb{23}$ products can always be eliminated to lead to an independent set.%
\footnote{Note we will not exploit relations like $\sqb{13}\sqb{24} = -\sqb{14}\sqb{23} s_{13} /s_{14}$ which would allow to keep one single such spinor structure but, being non local, are inconvenient when constructing non-factorizable contact terms.}
An application of this procedure to a five-point amplitude is discussed in \autoref{sec:sm-eft}.

To obtain more symmetric combinations in four-point amplitudes, one may however wish to perform the spinor structure reduction in a more tailored way.
For this purpose, we find it convenient to introduce the following shorthand notation:
\begin{align}
(lmn)\equiv\: \sqb{12}^l\sqb{34}^l\;\: \sqb{13}^m\sqb{24}^m\;\: \sqb{14}^n\sqb{23}^n\,.
\label{eq:lmn}
\end{align}
We also define the \emph{order} of a $(lmn)$ spinor structure as $\max\{l,m,n\}$.
In this notation, the Schouten identity writes $(100)-(010)+(001)=0$.
Each of these three terms carries helicity weight $1/2$ for each of the $1,2,3,4$ particles.
In various cases, minimal spinor structures compatible with little-group constraints can be expressed as a prefactor multiplying the various $(lmn)$ structure satisfying the $l+m+n=o$ constraint, for a positive integer $o$ (which is their maximal order).

Finally, minimal spinor structures can be complemented by scalar functions of momentum dot products $s_{ij}\equiv 2p_i\cdot p_j$.
In local non-factorizable amplitudes, they only appear in numerators and increase the mass dimension by at least two units.
All dot products however vanish in massless three-point amplitudes.
There are $(\nl-1)(\nl-2)/2$ available dot products in amplitudes with $\nl$ external legs, once one particle momentum is eliminated using momentum conservation.
One on-shell condition reduces this number to $\nl(\nl-3)/2$ and Gram determinant relations reduce it further to $3\nl-10$ (for $\nl>3$, see e.g.\ \cite{Duhr:2011}).
A systematic way to build a set of independent dot products based on the kinematic polynomial rings can be found in ref.~\cite{Henning:2017fpj}.

\section{Minimal operator dimension}
\label{sec:min_op_dim}

The dimension of an amplitude with $\nl$ external legs is $4-\nl$.
It receives contributions from the dimension of the spinor structure \dspin\ as well as from the coefficient of the operator that generated it, which has dimension $4-\dop$.
From the operator coefficient dimension, one can also extract a factor of the Planck mass for each external graviton, to reproduce the operator dimension assignment of ref.~\cite{Ruhdorfer:2019qmk}.
One thus have
\begin{equation}
\dop = \nl-\nt+\dspin
\end{equation}
where $\nt$ is the number of external Lorentz tensor legs (i.e.\ gravitons).
Determining the minimal operator dimension at which a given non-factorizable helicity amplitude arises thus requires a determination of the dimension of the minimal spinor structure allowed for that amplitude.

Forbidding spinor brackets in denominators, the dimension of the spinor structure yielding a given helicity amplitude is at least that of the spinors needed to generate the suitable helicity weights.
One spinor of dimension $1/2$ is required for each half unit of helicity.
The spinor structure required for a non-factorizable amplitude featuring external legs of helicities $h_i$ should thus have dimension at least equal to $\sum_i |h_i|$: $\dspin \ge \sum_i |h_i|$.
This inequality is saturated when the $\sum_i |2h_i|$ spinors required to generate the correct helicity weights are all contracted in bilinears.
This may however not always be possible.
Additional momentum insertions are otherwise required and further increase the dimension of the spinor structure leading to 
\begin{equation}
\dspin = \sum_i |h_i| + n_\text{mom.\ ins.}\,.
\end{equation}%
Momentum insertions%
\footnote{Note that a light-like momentum can always be decomposed as a product of spinors: $p_i \equiv p _i^\mu \sigma_\mu = i \ra\lb i$ and $\bar p_i\equiv p_i^\mu\bar\sigma_\mu = i]\langle i$ (suppressing spinor indices).
So momentum insertions can be written as bilinears such that e.g.\ $\la 231 \rb = \la 2 3 \ra \lb 31 \rb$.
For convenience, we however choose to not split momentum insertions in this form.}
allow to contract together angle and square brackets as in $\langle 1|p_2|3]\equiv \langle 123]$, or spinors of the same particle as in $[1|\bar{p}_2p_3|1]\equiv[1231]$.
In three-point amplitudes, such structures featuring additional momentum insertions however vanish due to momentum conservation and on-shell conditions.

To determine the number of required momentum insertions, one starts by considering separately particles carrying positive and negative helicities, which respectively demand square and angle brackets.
By Lorentz invariance, the total number of spinors is always even.
At least a momentum insertion is required when the number of spinor of either kind is odd.
This for instance occurs in the $(+1/2,+1/2,+1/2,-1/2,-1/2,-1/2)$ six-fermion amplitude,\footnote{We thank the authors of ref.~\cite{Li:2020gnx} for pointing out this particular case and amending our \autoref{eq:mom_ins} to also cover it.} in the analogous ten-fermion, or spin-$3/2$ cases.
So $n_\text{mom.\ ins.}\ge \big\{\sum_{h_i>0} 2h_i\big\} \mod 2 = \big\{\sum_{h_i<0} 2h_i\big\} \mod 2$.
Additional momentum insertions can moreover be required in amplitudes involving particles of unequal spins.
The crucial point is that, if the largest positive helicity exceeds the sum of all other positive ones, its square brackets cannot all be contracted in bilinears.
The same holds for the largest negative helicity.
In the example of a four-point amplitude with $(+2,+1,+1/2,+1/2)$ positive helicities, the $+2$ helicity requires four square brackets which can all be contracted with the four square brackets required for the other three particles.
In a $(+2,+1,+1/2,-1/2)$ helicity amplitude, however, particles of helicity $+1$ and $+1/2$ only provide three square brackets which is insufficient to contract all the ones required by the particle of helicity $+2$.
One momentum insertion is required to contract its fourth square bracket with the angle bracket needed for the particle of helicity $-1/2$.

In general, one remains with $\:2\max_{h_i>0}\{|2h_i|\} - \sum_{h_{i}>0} |2h_i|$ uncontracted square spinors and $\:2\max_{h_i<0}\{|2h_i|\} - \sum_{h_{i}<0} |2h_i|$ angle spinors, if those numbers are positive.
(The factor of two in front of the maximal helicity in the first term is required if the sum of the second term also includes it.)
One can then for instance form $\langle 123]$ structures, with one momentum insertion, for each available pair of square and angle brackets.
If they remain in unequal number, one would also need to form $\sqb{1231}$ or $\anb{1231}$ structures with two momentum insertions.
Forming preferentially trilinears or quadrilinears yields the same counting, as one can for instance split the product of a quadrilinear and a bilinear into two trilinears.
Overall, the number of required momentum insertions is equal to the maximum between the numbers of remaining square and angle brackets:
\begin{equation}
n_\text{mom.\ ins.} \ge 
\max\left[
	\begin{array}{c}
	\big\{\sum_{h_i>0} 2h_i\big\} \mod 2\\
	2\max\limits_{h_i>0}\{|2h_i|\} - \sum\limits_{h_{i}>0} |2h_i|\\
	2\max\limits_{h_i<0}\{|2h_i|\} - \sum\limits_{h_{i}<0} |2h_i|
	\end{array}
	\right]
\,.
\label{eq:mom_ins}
\end{equation}
A consequence of this formula is that amplitudes requiring a minimal number of momentum insertions that is odd involve fermions.

Our final expression for the minimal operator dimension at which a given non-factorizable amplitude is generated thus takes the form
\begin{multline}
\dop \ge  \nl - \nt \\
+ \!\sum_i|h_i|+ 
\max\left[
	\begin{array}{@{}c@{}}
	\big\{\sum_{h_i>0} 2h_i\big\} \mod 2\\
	2\max\limits_{h_i>0}\{|2h_i|\} - \sum\limits_{h_{i}>0} |2h_i|\\
	2\max\limits_{h_i<0}\{|2h_i|\} - \sum\limits_{h_{i}<0} |2h_i|
	\end{array}
	\right]\,,
\label{eq:dim}
\end{multline}
where again $\nl$ is the number of external legs, $\nt$ the number of external tensors (i.e.\ gravitons), and $h_i$ are the helicities of each leg.
This formula is only applicable for $\nl>3$.
If the number of required momentum insertions is non-vanishing, no local non-renormalizable three-point amplitude can be written down.

In the applications considered below, we find that the inequality of \autoref{eq:dim} can always be saturated, before other constraints such as gauge invariance or spin statistics are imposed.

\section{Operator enumerations}
\label{sec:enumerations}

We generically denote as $s,f,v,t$ massless particles of spin $0,1/2,1,2$ and indicate the sign of their helicity with a $\pm$ superscript.
Helicity amplitudes with more and more legs are successively considered.
For simplicity, we restrict ourselves to amplitudes of non-negative total helicity.
The others can be obtained by trading square for angle brackets, and vice versa (i.e.\ by flipping parity).
The counting rule of \autoref{eq:dim} is employed to determine the minimal dimension of operators giving rise to each of these helicity amplitudes.
Focusing on those generated by operators of dimension eight at most, one then solves the little-group constraints discussed in \autoref{sec:lg} to generate the corresponding minimal spinor structures.
Straightforward in most cases, the reduction of spinor structures to an independent set relies on momentum conservation, on-shell conditions, and Schouten identities.
Dot products of momenta can eventually be appended to minimal independent spinor structures.
Re-considering the SM diversity of fermions and gauge bosons, as well as non-trivial gauge transformation properties, is then required.
Bose and Fermi statistics also need to be imposed in the presence of identical fields.
An enumeration of independent operators is finally obtained.

\newcommand{\no}{\ding{55}}
\newcommand{\No}{}
\newcommand{\X}{\ding{55}}
\newcommand{\V}{\ding{51}}
\newcommand{\yes}{}
\begin{table*}\centering
\adjustbox{max width=\textwidth}{\begin{tabular}{@{}l@{\quad}c@{\quad}ccccc@{}}
mult.
	& min.\ dim.
	& helicity conf.
	& spinor structures
	& \rotatebox{90}{SM gauge}
	& \rotatebox{90}{spin stat.}
	& Hilbert series
\\[1mm]\hline\noalign{\vskip1mm}
3-pt
& dim-5
	& $t^+t^+s$
	& $\sqb{12}^4$
	& \no	&
\\[1mm]
& dim-6
	& $t^+t^+t^+$
	& $\sqb{12}^2\sqb{13}^2\sqb{23}^2$
	& \yes	&
	& $C_R^3$
\\
	&& $t^+t^+v^+$
	& $\sqb{12}^3\sqb{23}\sqb{13}$
	& \yes	&\X
\\
	&& $t^+v^+v^+$
	& $\sqb{12}^2\sqb{13}^2$
	& \yes	&
	& $(B_R^2,W_R^2,G_R^2)C_R$
\\[1mm]\hline\noalign{\vskip1mm}
4-pt
& dim-6
	& $\sym t^+\sym t^+ss$
	& $\sqb{12}^4$; $\sqb{12}^4s_{12}$
	& \yes	&
	& $HC_R^2H^\dagger$, $HD^2H^\dagger C_R^2$
\\[1mm]
& dim-7
	& $t^+t^+t^+ s$
	& $\sqb{12}^2\sqb{13}^2 \sqb{23}^2$
	& \no	&
\\
	&& $t^+t^+v^+s$
	& $\sqb{12}^3\sqb{13} \sqb{23}$
	& \no	& \X
\\
	&& $\sym t^+\sym t^+f^+f^+$
	& $\sqb{12}^4\sqb{34}$
	& \no	&
\\
	&& $t^+t^+f^-f^-$
	& $\sqb{12}^4\anb{34}$
	& \no	&
\\
	&& $t^+v^+v^+s$
	& $\sqb{12}^2\sqb{13}^2$
	& \no	&
\\
	&& $t^+v^+f^+f^+$
	& $\sqb{12}^2\sqb{13} \sqb{14}$
	& \no	&
\\[1mm]
& dim-8
	& $\sym t^+\sym t^+\sym t^+\sym t^+$
	& $\sqb{12}^4\sqb{34}^4\! +\! \sqb{13}^4\sqb{24}^4\! +\! \sqb{14}^4 \sqb{23}^4$
	& \yes &
	& $C_R^4$
\\
	&& $t^+t^+t^+v^+$
	& $\sqb{12}^3 \sqb{13}\sqb{23}\sqb{34}^2$,
	  $\sqb{12} \sqb{13}^3\sqb{23}\sqb{24}^2$,
	  $\sqb{12} \sqb{13}\sqb{23}^3\sqb{14}^2$
	& \yes	&\X
\\
	&& $t^+t^+t^-t^-$
	& $\sqb{12}^4 \anb{34}^4$
	& \yes	&
	& $C_R^2C_L^2$
\\
	&& $\sym t^+\sym t^+v^+v^+$
	& $\sqb{12}^4\sqb{34}^2$, $\sqb{12}^2\sqb{13}\sqb{14}\sqb{24}\sqb{23}$
	& \yes	&
	& $2(B_R^2,W_R^2,G_R^2)C_R^2$
\\
	&& $t^+t^+v^-v^-$
	& $\sqb{12}^4\anb{34}^2$
	& \yes	&
	& $(B_L^2,W_L^2,G_L^2)C_R^2$
\\
	&& $t^+t^+f^+f^-$
	& $\sqb{12}^4[3(1-2)4\rangle$
	& \yes	&\X
\\
	&& $t^+v^+\sym v^+\sym v^+$
	& $\sqb{12} \sqb{13} \sqb{14} (\sqb{13}\sqb{24}+\sqb{14}\sqb{23})$
	& \yes	&
	& $(W_R^2,G_R^2)B_RC_R$
\\
	&& $t^+v^+f^+f^-$
	& $\sqb{12}^2 \sqb{13} [124\rangle$
	& \yes	&
	& $(QQ^\dagger,uu^\dagger,dd^\dagger,LL^\dagger,ee^\dagger) DB_RC_R$,
	\\&&&&&& $(QQ^\dagger, LL^\dagger) DW_RC_R$,
	\\&&&&&& $(QQ^\dagger,uu^\dagger, dd^\dagger) DG_RC_R$
\\
	&& $t^+v^+ss$
	& $\sqb{12}^2 \sqb{1231}$
	& \yes	&
	& $(B_R,W_R)HH^\dagger D^2 C_R$
\\
	&& $t^+f^+f^+s$
	& $\sqb{12}\sqb{13}\sqb{1231}$
	& \yes	&
	& $(Q^\dagger u^\dagger H^\dagger,Q^\dagger d^\dagger H,L^\dagger e^\dagger H )D^2C_R$
\\[1mm]
& ...
\\[1mm]\hline\noalign{\vskip1mm}
5-pt
& dim-7
	& $t^+t^+sss$
	& $\sqb{12}^4$
	& \no	&
\\[1mm]
& dim-8
	& $t^+t^+t^+ss$
	& $\sqb{12}^2\sqb{13}^2\sqb{23}^2$
	& \yes	&
	& $HH^\dagger C_R^3$
\\
	&& $t^+t^+v^+ss$
	& $\sqb{12}^3\sqb{13} \sqb{23}$
	& \yes	&\X
\\
	&& $\sym t^+\sym t^+f^+f^+s$
	& $\sqb{12}^4 \sqb{34}$
	& \yes	&
	& $(Q^\dagger u^\dagger H^\dagger,Q^\dagger d^\dagger H,L^\dagger e^\dagger H )C_R^2$
\\
	&& $t^+t^+f^-f^-s$
	& $\sqb{12}^4 \anb{34}$
	& \yes	&
	& $(Q u H,Q d H^\dagger,L e H^\dagger )C_R^2$
\\
	&& $t^+v^+v^+ss$
	& $\sqb{12}^2\sqb{13}^2$
	& \yes	&
	& $(B_R^2,B_RW_R,W_R^2,G_R^2)HH^\dagger C_R$
\\
	&& $t^+v^+f^+f^+s$
	& $\sqb{12}^2\sqb{13}\sqb{14}$
	& \yes	&
	& $(Q^\dagger u^\dagger H^\dagger,Q^\dagger d^\dagger H,L^\dagger e^\dagger H )(B_R,W_R)C_R$
	\\&&&&&& $(Q^\dagger u^\dagger H^\dagger,Q^\dagger d^\dagger H )G_RC_R$
\\
	&& $t^+f^+f^+f^+f^+$
	& $\sqb{12}\sqb{13}\sqb{14}\sqb{15}$
	& \yes	&
	& $Q^\dagger Q^\dagger Q^\dagger L^\dagger{}C_R, d^\dagger e^\dagger u^\dagger {}^2C_R$,
	\\&...&&&&& $d^\dagger Q^\dagger {}^2u^\dagger C_R, e^\dagger L^\dagger Q^\dagger u^\dagger C_R$
\\[1mm]\hline\noalign{\vskip1mm}
6-pt
& dim-8
	& $t^+t^+ssss$
	& $\sqb{12}^4$
	& \yes	&
	& $H^2 H^\dagger{}^2C_R^2$
\\[1mm]
& ...
\\[1mm]\hline\noalign{\vskip1mm}
7-pt
& dim-9
	& $t^+t^+sssss$
	& $\sqb{12}^4$
	& \X&
\\[1mm]
& ...
\\[1mm]\hline
\end{tabular}}
\caption{Non-factorizable helicity amplitudes and spinor structures produced by operators of dimension up to eight, involving at least one massless Lorentz tensor, together with vectors, fermions and scalars.
A cross mark in the \emph{SM gauge} column indicates the incompatibility of the corresponding helicity amplitude with $SU(3)_C\times SU(2)_L\times U(1)_Y$ gauge invariance.
Symmetry under the exchange of the underlined particles is explicitly imposed in the spinor structures displayed.
A cross mark in the \emph{spin stat.} column indicates no structure compatible with Bose statistics can be constructed when the Lorentz tensors are identical.
The terms of the Hilbert series of ref.~\cite{Ruhdorfer:2019qmk} corresponding to each helicity amplitude are provided in the last column.
The $Q^\dagger Q^\dagger Q^\dagger L^\dagger C_R$ term vanishes with one single generation of quarks and does not appear explicitly in ref.~\cite{Ruhdorfer:2019qmk}.
}
\label{tab:amps}
\end{table*}

\renewcommand{\V}{}
\renewcommand{\X}{}
\begin{table*}\centering
\adjustbox{max height=.5\textheight}{\begin{tabular}{l@{\quad}c@{\quad}ccccc}
mult.
	& min.\ dim.
	& helicity conf.
	& spinor structures
\\[1mm]\hline\noalign{\vskip1mm}
3-pt
	& dim-3
		& $sss$
			& constant
				& \X
\\	& dim-4
		& $f^+f^+s$
			& $\sqb{12}$
				& \V
\\	& dim-5
		& $v^+v^+s$
			& $\sqb{12}^2$
				& \X
\\	&	& $v^+f^+f^+$
			& $\sqb{12}\sqb{13}$
				& \V
\\	& dim-6
		& $v^+v^+v^+$
			& $\sqb{12}\sqb{13}\sqb{23}$
\\[1mm]\hline\noalign{\vskip1mm}
4-pt
	& dim-4
		& $ssss$
			& constant;
			$s_{ij}$;
			$s_{ij}s_{kl}$
\\	& dim-5
		& $f^+f^+ss$
			& $\sqb{12}(s_{ij})$
				& \X
\\	& dim-6
		& $v^+v^+ss$
			& $\sqb{12}^2	(s_{ij})$
\\	&	& $v^+f^+f^+s$
			& $\sqb{12}\sqb{13}(s_{ij})$
\\	&	& $f^+f^+f^+f^+$
			& $\sqb{12}\sqb{34}	(s_{ij})$,
			$\sqb{13}\sqb{24}(s_{ij})$
\\	&	& $f^+f^+f^-f^-$
			& $\sqb{12}\anb{34}(s_{ij})$
\\	&	& $f^+f^-ss$
			& $[1(3-4)2\rangle (s_{ij})$
\\	& dim-7
		& $v^+v^+v^+s$
			& $\sqb{12}\sqb{13}\sqb{23}$
				& \X
\\	&	& $v^+v^+f^+f^+$
			& $\sqb{12}^2\sqb{34}, \sqb{12}(\sqb{14}\sqb{23}+\sqb{13}\sqb{24})$
				& \X
\\	&	& $v^+v^+f^-f^-$
			& $\sqb{12}^2\anb{34}$
				& \X
\\	&	& $v^+f^+f^-s$
			& $\sqb{12}[123\rangle$
				& \X
\\	&	& $v^+sss$
			& $\sqb{1231}$
				& \X
\\	&	& $f^+f^+f^+f^-$
			& $\sqb{12}[3(1\!-\!2)4\rangle$
				& \X
\\	& dim-8
		& $v^+ v^+v^+v^+$
			& $\sqb{12}^2\sqb{34}^2,\sqb{13}^2\sqb{24}^2,\sqb{14}^2\sqb{23}^2$ 
\\	&	& $v^+v^+v^-v^-$
			& $\sqb{12}^2\anb{34}^2$
\\	&	& $v^+v^+f^+f^-$
			& $\sqb{12}^2[3(1\!-\!2)4\rangle$
\\	&	& $v^+v^-f^+f^-$
			& $\sqb{13}\anb{24}[1(3\!-\!4)2\rangle$
\\	&	& $v^+v^-ss$
			& $[1(3-4)2\rangle^2$
\\	&	& $v^+f^-f^-s$
			& $\sqb{1231}\anb{23}$
\\	& dim-9
		& $v^+v^+v^-s$
			& $\sqb{12}^2 \anb{3123}$
				& \X
\\	&	& $v^+v^-f^+f^+$
			& $[34][1(3\!-\!4)2\rangle^2$
				& \X
\\	& dim-10
		& $v^+v^+v^+v^-$
			& $[12]^2[3(1-2)4\rangle^2$
\\[1mm]\hline\noalign{\vskip1mm}
5-pt
	& dim-5
		& $sssss$
			& constant; $s_{ij}$
				& \X
\\	& dim-6
		& $f^+f^+sss$
			& $[12](s_{ij});\sqb{1342}$
\\	& dim-7
		& $v^+v^+sss$
			& $\sqb{12}^2$
				& \X
\\	&	& $v^+f^+f^+ss$
			& $\sqb{12}\sqb{13}$
				& \X
\\	&	& $f^+f^+f^+f^+s$
			& $\sqb{12}\sqb{34},\sqb{13}\sqb{24}$
				& \X
\\	&	& $f^+f^+f^-f^-s$
			& $\sqb{12}\anb{34}$
				& \X
\\	&	& $f^+f^-sss$
			& $[132\rangle,[142\rangle$
				& \X
\\	& dim-8
		& $v^+v^+v^+ss$
			& $\sqb{12}\sqb{13}\sqb{23}$
\\	&	& $v^+v^+f^+f^+s$
			& $\sqb{12}^2\sqb{34}, \sqb{12}(\sqb{14}\sqb{23}+\sqb{13}\sqb{24})$
\\	&	& $v^+v^+f^-f^-s$
			& $\sqb{12}^2\anb{34}$
\\	&	& $v^+f^+f^+f^+f^+$
			& $[13][14][25],[13][15][34],[14][15][23]$
\\	&	& $v^+f^+f^+f^-f^-$
			& $\sqb{12}\sqb{13}\anb{45}$
\\	&	& $v^+f^+f^-ss$
			& $\sqb{12}[143\rangle,\sqb{12}[123\rangle$
\\	&	& $v^+ssss$
			& $\sqb{1341},\sqb{1241},\sqb{1231}$
\\	&	& $f^+f^+f^+f^-s$
			& $\sqb{13}[234\rangle,\sqb{12}[324\rangle,\sqb{12}[314\rangle$
\\	&$\cdots$
\\[1mm]\hline\noalign{\vskip1mm}
6-pt
	& dim-6
		& $ssssss$
			& constant, $s_{ij}$
\\	& dim-7
		& $f^+f^+ssss$
			& $\sqb{12}$
\\ 	& dim-8
		& $v^+v^+ssss$
			& $\sqb{12}^2$
\\	&	& $v^+f^+f^+sss$
			& $\sqb{12}\sqb{13}$
\\	&	& $f^+f^+f^+f^+ss$
			& $\sqb{12}\sqb{34},\sqb{13}\sqb{24}$
\\	&	& $f^+f^+f^-f^-ss$
			& $\sqb{12}\anb{34}$
\\	&	& $f^+f^-ssss$
			& $[142\rangle,[152\rangle,[132\rangle$
\\	& $\cdots$
\\[1mm]\hline\noalign{\vskip1mm}
7-pt
	& dim-7
		& $sssssss$
			& constant
\\	& dim-8
		& $f^+f^+sssss$
			& $\sqb{12}$
\\	& $\cdots$
\\[1mm]\hline\noalign{\vskip1mm}
8-pt
	& dim-8
		& $ssssssss$
			& constant
\\	& $\cdots$
\\[1mm]\hline
\end{tabular}}
\caption{Non-factorizable helicity amplitudes generated by operators of dimension eight at most, involving massless Lorentz vectors, fermions and scalars.
The non-local renormalizable three-point amplitudes arising from gauge couplings are excluded.
The corresponding independent spinor structures are listed in the last column.
No symmetrization is explicitly required but linear combinations of definite transformation properties are discussed in \autoref{sec:sm-eft}.
}
\label{tab:sm_amps}
\end{table*}

We consider below the case of operators involving at least a graviton and SM particles.
Additionally, for pure SM operators, we focus on the enumeration of independent spinor structures without accounting for gauge ones and momentum dot products.
For a single generation, $993$ independent dimension-eight operators would be generated in this SMEFT case~\cite{Henning:2015alf}.

\subsection{GRSMEFT}
\label{sec:gr-sm-eft}

The (positive) helicity amplitudes generated by operators of dimension eight at most and featuring at least a Lorentz tensor (i.e.\ graviton) are listed in \autoref{tab:amps}.

\subparagraph{Gauge symmetry}
The SM gauge symmetries translates into simple selection rules:
\begin{list}{--}{\leftmargin3.5mm \topsep0pt \itemsep0pt \parsep0pt}
\item forcing pairs same-helicity fermions to appear together with a Higgs field, or with another such pair,
\item forbidding odd numbers of Higgs fields in the absence of a same-helicity fermion pair,
\end{list}
They exclude all amplitudes corresponding to operators of odd dimensions in \autoref{tab:amps}.
Demanding that the colour, weak isospin and hypercharge of gauge bosons, fermions, and Higgs fields compensate each other is eventually required once the nature of $s,f,v,t$ is specified.

\subparagraph{Explicit reduction and spin statistics}

Let us reduce the spinor structures compatible with little-group constraints to independent sets, and impose the symmetry required by spin statistics under the exchange of gravitons carrying identical helicities.

As discussed in \autoref{sec:lg}, three-point amplitudes have a unique form in terms of powers of $[12],[23],[13]$ spinor bilinears (for positive total helicity).
No spinor structure reduction thus needs to be carried out in this case.
The $t^+t^+v^+$ amplitude has a $\sqb{12}^3\sqb{23}\sqb{13}$ spinor structure antisymmetric under the exchange of the two Lorentz tensors and is therefore forbidden when these represent identical gravitons.
The same conclusion also holds for the $t^+t^+v^+s(s)$ spinor structures.

One scalar can always be ignored in the construction of higher-point amplitudes.
It only manifests itself in the form of insertions of its momentum which can be traded for others using momentum conservation.
One single minimal spinor structure, corresponding to the three-point amplitude solution, is thus found for the following four-point amplitude: $t^+t^+ss$, $t^+t^+t^+s$, $t^+t^+v^+s$, $t^+v^+v^+s$, $t^+v^+ss$, $t^+f^+f^+s$.
The higher-point amplitudes, like $t^+t^+sss$, $t^+t^+t^+ss$, $t^+t^+v^+ss$ and $t^+v^+v^+ss$, featuring at most three particles of non-vanishing helicities and requiring no momentum insertion also have one single independent minimal spinor structure matching that of the corresponding three-point amplitude.

One single minimal spinor structure is also found for the $t^+t^+f^-f^-(s)$, $t^+t^+t^-t^-$ and $t^+t^+v^-v^-(s)$ amplitudes involving particles of both positive and negative helicities but no momentum insertion.
Their square and angle brackets are contracted separately.
The $t^+t^+f^+f^-$ and $t^+v^+f^+f^-$ amplitudes which involve one single particle of negative helicity both have minimal spinor structures related by momentum conservation:
$\sqb{12}^3 \times \left( \sqb{12}[314\rangle = -\sqb{12}[324\rangle = \sqb{13}[234\rangle\right)$, and $\sqb{12}\sqb{13} \times \left( [134\rangle = -[124\rangle \right)$, respectively.
The first one can thus always be expressed as $[12]^4[3(1-2)4\rangle$ and is seen to be antisymmetric under the exchange of the two Lorentz tensors, which is incompatible with Bose statistics if they are identical gravitons.

Among amplitudes featuring only particles of positive helicities, the $t^+v^+f^+f^+(s)$ and $t^+f^+f^+f^+f^+$ ones only have one single minimal spinor structure: $\sqb{12}^2\sqb{13}\sqb{14}$ and $\sqb{12}\sqb{13}\sqb{14}\sqb{15}$, respectively.
No reduction is therefore needed in these cases.
The latter spinor structure is symmetric under the exchange of any pair of fermion.
After combination with colour and $SU(2)_L$ structures, this will imply that the $C_RQ^\dagger Q^\dagger Q^\dagger L^\dagger$ operator (in the notation of ref.~\cite{Ruhdorfer:2019qmk}) vanishes for a single generation.

The $t^+t^+f^+f^+(s)$ amplitudes have three minimal structures compatible with little-group constraints:
$\sqb{12}^3\sqb{13}\sqb{24}$,
$\sqb{12}^3\sqb{14}\sqb{23}$,
$\sqb{12}^4\sqb{34}$.
They are however related by the Schouten identity: $\sqb{12}^3\{(100)-(010)+(001)\}=0$ in the $(lmn)$ notation of \autoref{eq:lmn}.
Choosing structures of definite transformation properties under the $1\leftrightarrow 2$ exchange, one can for instance use $\sqb{12}^4\sqb{34}$ (even) and $\sqb{12}^3(\sqb{13}\sqb{24}+\sqb{14}\sqb{23})$ (odd).
When the two tensors represent identical gravitons, Bose statistics forbids the latter.
One is thus left with the former as single independent spinor structure.  

Similarly, the $t^+v^+v^+v^+$ amplitude has three minimal structures compatible with little-group constraints:
$\sqb{12}^2\sqb{34}\sqb{13}\sqb{14}$,
$\sqb{12}\sqb{13}^2\sqb{24}\sqb{14}$,
$\sqb{12}\sqb{13}\sqb{14}^2\sqb{23}$.
They are also related by the Schouten identity, $\sqb{12}\sqb{13}\sqb{14}\{(100)-(010)+(001)\}=0$, and one can construct combinations that are either symmetric or antisymmetric under the exchange of the last two Lorentz vectors (labelled $3$ and $4$):
$\sqb{12}^2\sqb{34}\sqb{13}\sqb{14}$ (odd) and
$\sqb{12}^2\sqb{13}\sqb{14}(\sqb{13}\sqb{24}+\sqb{14}\sqb{23})$ (even).
Due to gauge invariance in the SM, the $t^+v^+v^+v^+$ amplitude necessarily involves two identical vectors and must therefore be symmetric under their exchange.
One therefore remains with the latter spinor structure as single allowed one.

There are six minimal spinor structures satisfying little-group covariance for the $t^+t^+v^+v^+$ amplitude.
In the $(lmn)$ notation introduced in \autoref{eq:lmn}, they are $\sqb{12}^2$ times $(200)$, $(020)$, $(002)$, $(110)$, $(101)$, or $(011)$.
The Schouten identity however allows us to express the structures of order two in terms of that of order one (and vice versa), as in $(200)=(110)-(101)$.
One thus remains with three independent spinor structures which can be chosen to have definite symmetry under the exchange of the two tensors.
Given that the $1\leftrightarrow2$ permutation sends $(100)$ to minus itself, and $(010)$ to $(001)$, the two symmetric and singe antisymmetric structures are respectively $\sqb{12}^2(011)$, $\sqb{12}^2\{(110)-(101)\}=\sqb{12}^2(200)$ and $\sqb{12}^2\{(110)+(101)\}$.
When the two tensors represent identical gravitons, only the symmetric structures survive: $\sqb{12}^2\sqb{13}\sqb{14}\sqb{24}\sqb{23}$ and $\sqb{12}^4\sqb{34}^2$.
They are also symmetric under the exchange of the two vectors.

There are six minimal spinor structures satisfying little-group covariance for the $t^+t^+t^+v^+$ amplitude.
They are $\sqb{12}\sqb{13}\sqb{23}$ times $(200)$, $(020)$, $(002)$, $(110)$, $(101)$, or $(011)$.
Squaring Schouten identities like $(100)=(010)-(001)$, one obtains that order-one structures can be expressed as combinations of order-two ones, e.g.\ $(011)=[(020)+(002)-(200)]/2$.
One can thus retain the latter and get three independent spinor structures: $\sqb{12}\sqb{13}\sqb{23}$ times $(200)$, $(020)$, or $(002)$.
Each of those three structures is antisymmetric under the exchange of a pair of the three tensors: under $1\leftrightarrow2$ for the first one, under $1\leftrightarrow3$ for the second one, and under $2\leftrightarrow3$ for the third one.
When all three represent identical gravitons, no structure ends up compatible with Bose statistics.

Finally, the $t^+t^+t^+t^+$ amplitude has fifteen minimal spinor structures satisfying little-group covariance.
Employing again the $(lmn)$ notation of \autoref{eq:lmn}, they are:
\begin{center}
\begin{tabular}{ll}
order four:	& $(400),(040),(004)$,\\
order three:	& $(310),(301),(031),(130),(013),(031)$,\\
order two:	& $(220),(022),(202),(211),(121),(112)$.
\end{tabular}
\end{center}
All structures of order three can be reduced to structures of order two using the Schouten identity, as in $(310)=(220)-(211)$.
Half of the structures of order two can also be eliminated using identities like $(211) = (200)\times(011) = (200) [(020)+(002)-(200)]/2$.
There remains six $(400),(040),(004),(220),(022),(202)$ spinor structures.
One of their linear combinations however vanishes since $(400)+(040)+(004)=2\{(220)+(202)+(022)\}$.
This relation can be obtained by squaring the $(011) = [(020)+(002)-(200)]/2$ equality already used above.
One thus finally obtains five independent spinor structures.
The same counting is for instance also obtained by eliminating all $(lmn)$ structures with $l>0$ by using the $(100)=(010)-(001)$ Schouten identity recursively, following the systematic procedure described in \autoref{sec:lg}.
When all four tensors represent identical gravitons, one can form two $(400)+(040)+(004)$, $(220)+(022)+(202)$ fully symmetric combinations.
They are however proportional to each other, as just demonstrated.
There is thus one single independent minimal spinor structure for the amplitude with four identical tensors of same helicity.

\subparagraph{Momentum dot products}

Dot products of momenta can be added to minimal spinor structures and would increase the corresponding operator dimension by at least two units.
Limiting ourselves to operators of dimension eight at most, one thus only needs to consider adding such $s_{ij}$'s to structures already leading to operators of dimension six at most.
In massless three-point amplitudes, all dot products of momenta vanish.
Examining \autoref{tab:amps}, there is thus only the dimension-six four-point $t^+t^+ss$ amplitude with minimal $[12]^4$ spinor structure to which one single momentum dot product can be appended to reach dimension eight.
In such a massless four-point amplitude, there are three possible products summing to zero: $s_{12}+s_{13}+s_{23}=0$.
One can thus for instance form two linear combinations of definite symmetry transformation properties under the exchange of the two tensors: $s_{12}$ (even) and $s_{13}-s_{23}$ (odd).
When they represent identical gravitons, Bose statistics only permits the former one (given that the spinor structure is already symmetric).

\subsection{SMEFT}
\label{sec:sm-eft}

Considering the enumeration of pure SMEFT operators, we focus on their spinor structures and do not account for gauge and flavour transformation properties.
Factors of momentum dot products are not explicitly specified either.
We still only construct explicitly amplitudes having non-negative total helicity, as the others can be obtained by exchanging square and angle brackets.
The helicity amplitudes generated by operators of dimension eight at most are displayed in \autoref{tab:sm_amps}.
For many helicity amplitudes, little-group constraints only allow for one single spinor structure corresponding to an operator of dimension eight at most.
Using momentum conservation and Schouten identities, the reduction of the obtained minimal spinor structures is straightforward for $f^+f^+f^+f^+(s)(s)$, $v^+v^+f^+f^+(s)$, $f^+f^+f^+f^-$, $v^+v^+f^+f^-$, $v^+v^-f^+f^-$, $f^+f^+ss(s)$ amplitudes.
We thus only discuss below the reductions of spinor structures for the $v^+v^+v^+v^+$ and $v^+f^+f^+f^+f^+$ amplitudes.
Note that the $f^+f^+sss$ five-point amplitude receives not only a $[12]$ contribution at dimension six, but also a $[1342]$ one at dimension eight.
We have not found other helicity amplitudes receiving contributions from spinor structures of different dimensions.

In the $(lmn)$ notation introduced in \autoref{eq:lmn}, the minimal spinor structures compatible with little-group constraints for the $v^+v^+v^+v^+$ amplitude are $(200),(020),(002)$ of order two and $(110),(101),(011)$ of order one.
As seen before, squaring Schouten identities leads to equalities such as $(011)=[(020)+(002)-(200)]/2$ which allow us to eliminate order-one structures in favour of order-two ones.
One is therefore left with three independent structures.
Imposing symmetry under the exchange of the first two vectors, one remains with the $(200)$ and $(020)+(002)$ structures.
Incidentally, these are simultaneously symmetric under the exchanges of the last two vectors.
The single structure that is symmetric under the exchange of all four vectors is $(200)+(020)+(002)$.

The six minimal structures compatible with little-group constraints for the $v^+f^+f^+f^+f^+$ amplitude are of the $[1i][1j][kl]$ form with $i\ne j\ne k\ne l$ ranging from two to five.
As discussed in \autoref{sec:lg}, there are five ($\nl!/4!(\nl-4)!$) elementary Schouten identities in a five-point amplitude.
A different spinor does not appear in each of them.
They can for instance be taken as
\begin{equation}
\begin{aligned}{}
[23][45]&=[24][35]-[25][34] 	&\text{(not involving $[1$)},\\
[13][45]&=[14][35]-[15][34]	&\text{(not involving $[2$)},\\
[12][45]&=[14][25]-[15][24]	&\text{(not involving $[3$)},\\
[12][35]&=[13][25]-[15][23]	&\text{(not involving $[4$)},\\
[12][34]&=[13][24]-[14][23]	&\text{(not involving $[5$)}.
\end{aligned}
\end{equation}
The first equality is of no use here since all minimal $v^+f^+f^+f^+f^+$ spinor structures involve the $[1$ spinor.
The left-hand side of the second equality does not appear either.
Eliminating all $[12][45]$, $[12][35]$, $[12][34]$ bilinear products with the last three equalities leads to three independent spinor structures: $[13][14][25]$, $[13][15][34]$, $[14][15][23]$.
Each of them is symmetric under the exchange of a pair of fermion and antisymmetric under the exchange of the other.

\subparagraph{Four-point amplitudes beyond dimension eight}

Focusing on four-point amplitudes, it is remarkable that only three helicity amplitudes are not generated by operators of dimension eight or lower.
They are $v^+v^+v^-s$, $v^+v^-f^+f^+$, and $v^+v^+v^+v^-$.
Let us also reduce their minimal spinor structures to independent sets.

Only $[12]^2\anb{3123}$ is available for the $v^+v^+v^-s$ amplitude, and is antisymmetric under the exchange of the two vectors of positive helicity.

Five minimal structures are compatible with little-group constraints for the $v^+v^-f^+f^+$ amplitude: $[14][312\rangle[142\rangle$, $[13][412\rangle[132\rangle$, $[34][132\rangle[142\rangle$, $[34][142\rangle^2$, $[34][132\rangle^2$.
They can however all be reduced to a single one, for instance the $[34][1(3-4)2\rangle^2$ one, using just momentum conservation and on-shell conditions.
It is antisymmetric under the exchange of the two fermions.

Six minimal spinor structures are compatible with little-group constraints for the four-vector $v^+v^+v^+v^-$ amplitude: $[12][13][234\rangle[324\rangle$, $[12][13][234\rangle[314\rangle$, $[12]^2\linebreak[1][314\rangle\linebreak[1] [324\rangle$, $[23]^2\linebreak[1][134\rangle^2$, $[12]^2[324\rangle^2$, $[12]^2[314\rangle^2$.
With momentum conservation and on-shell conditions, they can be reduced to a single independent structure: $[12]^2[3(1-2)4\rangle^2$ which is symmetric under the exchange of the first two vectors.

We also verified that no independent higher-dimensional spinor structure contributes to any four-point helicity amplitudes, besides the ones displayed in \autoref{tab:sm_amps}.
For this purpose, besides the systematic treatment of Schouten identities already discussed, we introduced an algorithmic procedure to account for momentum conservation.
Any $[ij]\anb{jk}$ product is replaced by a $[il]\anb{lk}$ one for $l\ne i,j,k$, iteratively for $j=1,2,3$.
Such a procedure could in principle be generalized to higher-point amplitudes.
The sum in the $[ij]\anb{jk}=-\sum_{l\ne i,j,k} [il]\anb{lk}$ replacement then however contains more than a single term.

\section{Conclusions}

We advanced the program of systematic non-factorizable amplitude construction, establishing a simple formula for the minimal dimension of operators generating any massless helicity amplitude.
Solving little-group constraints then yielded minimal spinor structures, which could subsequently be reduced to independent sets using momentum conservation, on-shell conditions, and Schouten identities.
A systematic procedure to apply Schouten identities was presented.
Dressing these minimal spinor structures with dot products of momenta led to the independent non-factorizable Lorentz structures.
The enumeration of operators of dimension eight at most was provided for operators involving particles of spins $0,1/2,1,2$.
Simple selection rules were sufficient to impose the constraints due to standard-model gauge symmetry, and symmetrization of the Lorentz structures under particle permutations provided compliance with Bose statistics for gravitons of same helicity and identical vector bosons.
The minimal spinor structures generated by operators of dimension eight at most were also provided for the pure standard-model case, featuring only particles of spins $0,1/2,1$.
The combination of these with the full richness of standard-model gauge and flavour structures was not attempted.
Such a combination would be interesting to perform in the future.
Tools and techniques developed for operator constructions could then be exploited~\cite{Gripaios:2018zrz, Criado:2019ugp, Hays:2018zze, Fonseca:2019yya}.
The systematic construction of massive non-factorizable amplitudes is also a task remaining to be performed.
A counting rule for the number of independent minimal spinor structures and an algorithmic reduction procedure for amplitudes with more than four legs would also be useful to derive in the future.

\subparagraph{Note added}
Reference~\cite{Falkowski:2019zdo} which appeared on the day of our submission to the arXiv employed momentum twistors to trivialize the constraint arising from momentum conservation, in the enumeration of operators from on-shell amplitudes.

\section*{Acknowledgements}
We would like to thank Teppei Kitahara, Teng Ma, Yael Shadmi and Yaniv Weiss for discussions and useful comments on our manuscript.
C.S.M.\ thanks the Technion for its hospitality while part of this work was carried out.
This work was initiated at Aspen Center for Physics which is supported by National Science Foundation grant PHY-1607611.
The work of G.D.\ is supported in part at the Technion by a fellowship from the Lady Davis Foundation.
The work of C.S.M.\ is supported by the Alexander von Humboldt Foundation, in the framework of the Sofja Kovalevskaja Award 2016, endowed by the German Federal Ministry of Education and Research and also supported by  the  Cluster  of  Excellence  ``Precision  Physics,  Fundamental Interactions, and Structure of Matter'' (PRISMA$^+$ EXC 2118/1) funded by the German Research Foundation (DFG) within the German Excellence Strategy (Project ID 39083149).

\bibliographystyle{apsrev4-1_title}
\bibliography{refs}

\begin{thebibliography}{30}%
\makeatletter
\providecommand \@ifxundefined [1]{%
 \@ifx{#1\undefined}
}%
\providecommand \@ifnum [1]{%
 \ifnum #1\expandafter \@firstoftwo
 \else \expandafter \@secondoftwo
 \fi
}%
\providecommand \@ifx [1]{%
 \ifx #1\expandafter \@firstoftwo
 \else \expandafter \@secondoftwo
 \fi
}%
\providecommand \natexlab [1]{#1}%
\providecommand \enquote  [1]{``#1''}%
\providecommand \bibnamefont  [1]{#1}%
\providecommand \bibfnamefont [1]{#1}%
\providecommand \citenamefont [1]{#1}%
\providecommand \href@noop [0]{\@secondoftwo}%
\providecommand \href [0]{\begingroup \@sanitize@url \@href}%
\providecommand \@href[1]{\@@startlink{#1}\@@href}%
\providecommand \@@href[1]{\endgroup#1\@@endlink}%
\providecommand \@sanitize@url [0]{\catcode `\\12\catcode `\$12\catcode
  `\&12\catcode `\#12\catcode `\^12\catcode `\_12\catcode `\%12\relax}%
\providecommand \@@startlink[1]{}%
\providecommand \@@endlink[0]{}%
\providecommand \url  [0]{\begingroup\@sanitize@url \@url }%
\providecommand \@url [1]{\endgroup\@href {#1}{\urlprefix }}%
\providecommand \urlprefix  [0]{URL }%
\providecommand \Eprint [0]{\href }%
\providecommand \doibase [0]{http://dx.doi.org/}%
\providecommand \selectlanguage [0]{\@gobble}%
\providecommand \bibinfo [0]{\@secondoftwo}%
\providecommand \bibfield [0]{\@secondoftwo}%
\providecommand \translation [1]{[#1]}%
\providecommand \BibitemOpen [0]{}%
\providecommand \bibitemStop [0]{}%
\providecommand \bibitemNoStop [0]{.\EOS\space}%
\providecommand \EOS [0]{\spacefactor3000\relax}%
\providecommand \BibitemShut  [1]{\csname bibitem#1\endcsname}%
\let\auto@bib@innerbib\@empty
\bibitem [{\citenamefont{Cheung} \emph {et\,al.}(2017)\citenamefont{Cheung},
  \citenamefont{Kampf}, \citenamefont{Novotny}, \citenamefont{Shen}, and
  \citenamefont{Trnka}}]{Cheung:2016drk}%
  \BibitemOpen
  \bibfield{author}{\bibinfo{author}{\bibfnamefont{C.}\,\bibnamefont{Cheung}},
  \bibinfo{author}{\bibfnamefont{K.}\,\bibnamefont{Kampf}},
  \bibinfo{author}{\bibfnamefont{J.}\,\bibnamefont{Novotny}},
  \bibinfo{author}{\bibfnamefont{C.-H.} \bibnamefont{Shen}},  and
  \bibinfo{author}{\bibfnamefont{J.}\,\bibnamefont{Trnka}},
  }\bibfield{title}{\emph {\bibinfo{title}{{A Periodic Table of Effective Field
  Theories}}}, }\href {\doibase 10.1007/JHEP02(2017)020}
  {\bibfield{journal}{\bibinfo{journal}{JHEP}\,}\textbf{\bibinfo{volume}{02}}\,(\bibinfo{year}{2017})\,\bibinfo{pages}{020}},
  \Eprint {http://arxiv.org/abs/1611.03137}{arXiv:1611.03137
  [hep-th]}\BibitemShut {NoStop}%
\bibitem [{\citenamefont{Cheung} \emph {et\,al.}(2018)\citenamefont{Cheung},
  \citenamefont{Kampf}, \citenamefont{Novotny}, \citenamefont{Shen},
  \citenamefont{Trnka}, and \citenamefont{Wen}}]{Cheung:2018oki}%
  \BibitemOpen
  \bibfield{author}{\bibinfo{author}{\bibfnamefont{C.}\,\bibnamefont{Cheung}},
  \bibinfo{author}{\bibfnamefont{K.}\,\bibnamefont{Kampf}},
  \bibinfo{author}{\bibfnamefont{J.}\,\bibnamefont{Novotny}},
  \bibinfo{author}{\bibfnamefont{C.-H.} \bibnamefont{Shen}},
  \bibinfo{author}{\bibfnamefont{J.}\,\bibnamefont{Trnka}},  and
  \bibinfo{author}{\bibfnamefont{C.}\,\bibnamefont{Wen}},
  }\bibfield{title}{\emph {\bibinfo{title}{{Vector Effective Field Theories
  from Soft Limits}}}, }\href {\doibase 10.1103/PhysRevLett.120.261602}
  {\bibfield{journal}{\bibinfo{journal}{Phys. Rev.
  Lett.}\,}\textbf{\bibinfo{volume}{120}}\,(\bibinfo{year}{2018})\,\bibinfo{pages}{261602}},
  \Eprint {http://arxiv.org/abs/1801.01496}{arXiv:1801.01496
  [hep-th]}\BibitemShut {NoStop}%
\bibitem [{\citenamefont{Elvang} \emph {et\,al.}(2019)\citenamefont{Elvang},
  \citenamefont{Hadjiantonis}, \citenamefont{Jones}, and
  \citenamefont{Paranjape}}]{Elvang:2018dco}%
  \BibitemOpen
  \bibfield{author}{\bibinfo{author}{\bibfnamefont{H.}\,\bibnamefont{Elvang}},
  \bibinfo{author}{\bibfnamefont{M.}\,\bibnamefont{Hadjiantonis}},
  \bibinfo{author}{\bibfnamefont{C.~R.~T.} \bibnamefont{Jones}},  and
  \bibinfo{author}{\bibfnamefont{S.}\,\bibnamefont{Paranjape}},
  }\bibfield{title}{\emph {\bibinfo{title}{{Soft Bootstrap and
  Supersymmetry}}}, }\href {\doibase 10.1007/JHEP01(2019)195}
  {\bibfield{journal}{\bibinfo{journal}{JHEP}\,}\textbf{\bibinfo{volume}{01}}\,(\bibinfo{year}{2019})\,\bibinfo{pages}{195}},
  \Eprint {http://arxiv.org/abs/1806.06079}{arXiv:1806.06079
  [hep-th]}\BibitemShut {NoStop}%
\bibitem [{\citenamefont{Low} and \citenamefont{Yin}(2019)}]{Low:2019ynd}%
  \BibitemOpen
  \bibfield{author}{\bibinfo{author}{\bibfnamefont{I.}\,\bibnamefont{Low}} and
  \bibinfo{author}{\bibfnamefont{Z.}\,\bibnamefont{Yin}},
  }\bibfield{title}{\emph {\bibinfo{title}{{Soft Bootstrap and Effective Field
  Theories}}}, }\href {\doibase 10.1007/JHEP11(2019)078}
  {\bibfield{journal}{\bibinfo{journal}{JHEP}\,}\textbf{\bibinfo{volume}{11}}\,(\bibinfo{year}{2019})\,\bibinfo{pages}{078}},
  \Eprint {http://arxiv.org/abs/1904.12859}{arXiv:1904.12859
  [hep-th]}\BibitemShut {NoStop}%
\bibitem [{\citenamefont{Cheung} and
  \citenamefont{Shen}(2015)}]{Cheung:2015aba}%
  \BibitemOpen
  \bibfield{author}{\bibinfo{author}{\bibfnamefont{C.}\,\bibnamefont{Cheung}}
  and \bibinfo{author}{\bibfnamefont{C.-H.} \bibnamefont{Shen}},
  }\bibfield{title}{\emph {\bibinfo{title}{{Nonrenormalization Theorems without
  Supersymmetry}}}, }\href {\doibase 10.1103/PhysRevLett.115.071601}
  {\bibfield{journal}{\bibinfo{journal}{Phys. Rev.
  Lett.}\,}\textbf{\bibinfo{volume}{115}}\,(\bibinfo{year}{2015})\,\bibinfo{pages}{071601}},
  \Eprint {http://arxiv.org/abs/1505.01844}{arXiv:1505.01844
  [hep-ph]}\BibitemShut {NoStop}%
\bibitem [{\citenamefont{Bern} \emph {et\,al.}(2020)\citenamefont{Bern},
  \citenamefont{Parra-Martinez}, and \citenamefont{Sawyer}}]{Bern:2019wie}%
  \BibitemOpen
  \bibfield{author}{\bibinfo{author}{\bibfnamefont{Z.}\,\bibnamefont{Bern}},
  \bibinfo{author}{\bibfnamefont{J.}\,\bibnamefont{Parra-Martinez}},  and
  \bibinfo{author}{\bibfnamefont{E.}\,\bibnamefont{Sawyer}},
  }\bibfield{title}{\emph {\bibinfo{title}{{Nonrenormalization and Operator
  Mixing via On-Shell Methods}}}, }\href {\doibase
  10.1103/PhysRevLett.124.051601} {\bibfield{journal}{\bibinfo{journal}{Phys.
  Rev.
  Lett.}\,}\textbf{\bibinfo{volume}{124}}\,(\bibinfo{year}{2020})\,\bibinfo{pages}{051601}},
  \Eprint {http://arxiv.org/abs/1910.05831}{arXiv:1910.05831
  [hep-ph]}\BibitemShut {NoStop}%
\bibitem [{\citenamefont{Azatov} \emph {et\,al.}(2017)\citenamefont{Azatov},
  \citenamefont{Contino}, \citenamefont{Machado}, and
  \citenamefont{Riva}}]{Azatov:2016sqh}%
  \BibitemOpen
  \bibfield{author}{\bibinfo{author}{\bibfnamefont{A.}\,\bibnamefont{Azatov}},
  \bibinfo{author}{\bibfnamefont{R.}\,\bibnamefont{Contino}},
  \bibinfo{author}{\bibfnamefont{C.~S.} \bibnamefont{Machado}},  and
  \bibinfo{author}{\bibfnamefont{F.}\,\bibnamefont{Riva}},
  }\bibfield{title}{\emph {\bibinfo{title}{{Helicity selection rules and
  noninterference for BSM amplitudes}}}, }\href {\doibase
  10.1103/PhysRevD.95.065014} {\bibfield{journal}{\bibinfo{journal}{Phys.
  Rev.}\,}\textbf{\bibinfo{volume}{D95}}\,(\bibinfo{year}{2017})\,\bibinfo{pages}{065014}},
  \Eprint {http://arxiv.org/abs/1607.05236}{arXiv:1607.05236
  [hep-ph]}\BibitemShut {NoStop}%
\bibitem [{\citenamefont{Henning} \emph
  {et\,al.}(2017{\natexlab{a}})\citenamefont{Henning}, \citenamefont{Lu},
  \citenamefont{Melia}, and \citenamefont{Murayama}}]{Henning:2015alf}%
  \BibitemOpen
  \bibfield{author}{\bibinfo{author}{\bibfnamefont{B.}\,\bibnamefont{Henning}},
  \bibinfo{author}{\bibfnamefont{X.}\,\bibnamefont{Lu}},
  \bibinfo{author}{\bibfnamefont{T.}\,\bibnamefont{Melia}},  and
  \bibinfo{author}{\bibfnamefont{H.}\,\bibnamefont{Murayama}},
  }\bibfield{title}{\emph {\bibinfo{title}{{2, 84, 30, 993, 560, 15456, 11962,
  261485, ...: Higher dimension operators in the SM EFT}}}, }\href {\doibase
  10.1007/JHEP08(2017)016}
  {\bibfield{journal}{\bibinfo{journal}{JHEP}\,}\textbf{\bibinfo{volume}{08}}\,(\bibinfo{year}{2017}{\natexlab{a}})\,\bibinfo{pages}{016}},
  \Eprint {http://arxiv.org/abs/1512.03433}{arXiv:1512.03433
  [hep-ph]}\BibitemShut {NoStop}%
\bibitem [{\citenamefont{Henning} \emph {et\,al.}(2016)\citenamefont{Henning},
  \citenamefont{Lu}, \citenamefont{Melia}, and
  \citenamefont{Murayama}}]{Henning:2015daa}%
  \BibitemOpen
  \bibfield{author}{\bibinfo{author}{\bibfnamefont{B.}\,\bibnamefont{Henning}},
  \bibinfo{author}{\bibfnamefont{X.}\,\bibnamefont{Lu}},
  \bibinfo{author}{\bibfnamefont{T.}\,\bibnamefont{Melia}},  and
  \bibinfo{author}{\bibfnamefont{H.}\,\bibnamefont{Murayama}},
  }\bibfield{title}{\emph {\bibinfo{title}{{Hilbert series and operator bases
  with derivatives in effective field theories}}}, }\href {\doibase
  10.1007/s00220-015-2518-2} {\bibfield{journal}{\bibinfo{journal}{Commun.
  Math.
  Phys.}\,}\textbf{\bibinfo{volume}{347}}\,(\bibinfo{year}{2016})\,\bibinfo{pages}{363}},
  \Eprint {http://arxiv.org/abs/1507.07240}{arXiv:1507.07240
  [hep-th]}\BibitemShut {NoStop}%
\bibitem [{\citenamefont{Henning} and
  \citenamefont{Melia}(2019)}]{Henning:2019enq}%
  \BibitemOpen
  \bibfield{author}{\bibinfo{author}{\bibfnamefont{B.}\,\bibnamefont{Henning}}
  and \bibinfo{author}{\bibfnamefont{T.}\,\bibnamefont{Melia}},
  }\bibfield{title}{\emph {\bibinfo{title}{{Constructing effective field
  theories via their harmonics}}}, }\href {\doibase
  10.1103/PhysRevD.100.016015} {\bibfield{journal}{\bibinfo{journal}{Phys.
  Rev.}\,}\textbf{\bibinfo{volume}{D100}}\,(\bibinfo{year}{2019})\,\bibinfo{pages}{016015}},
  \Eprint {http://arxiv.org/abs/1902.06754}{arXiv:1902.06754
  [hep-ph]}\BibitemShut {NoStop}%
\bibitem [{\citenamefont{Shadmi} and
  \citenamefont{Weiss}(2019)}]{Shadmi:2018xan}%
  \BibitemOpen
  \bibfield{author}{\bibinfo{author}{\bibfnamefont{Y.}\,\bibnamefont{Shadmi}}
  and \bibinfo{author}{\bibfnamefont{Y.}\,\bibnamefont{Weiss}},
  }\bibfield{title}{\emph {\bibinfo{title}{{Effective Field Theory Amplitudes
  the On-Shell Way: Scalar and Vector Couplings to Gluons}}}, }\href {\doibase
  10.1007/JHEP02(2019)165}
  {\bibfield{journal}{\bibinfo{journal}{JHEP}\,}\textbf{\bibinfo{volume}{02}}\,(\bibinfo{year}{2019})\,\bibinfo{pages}{165}},
  \Eprint {http://arxiv.org/abs/1809.09644}{arXiv:1809.09644
  [hep-ph]}\BibitemShut {NoStop}%
\bibitem [{\citenamefont{Ma} \emph {et\,al.}()\citenamefont{Ma},
  \citenamefont{Shu}, and \citenamefont{Xiao}}]{Ma:2019gtx}%
  \BibitemOpen
  \bibfield{author}{\bibinfo{author}{\bibfnamefont{T.}\,\bibnamefont{Ma}},
  \bibinfo{author}{\bibfnamefont{J.}\,\bibnamefont{Shu}},  and
  \bibinfo{author}{\bibfnamefont{M.-L.} \bibnamefont{Xiao}},
  }\bibfield{title}{\emph {\bibinfo{title}{{Standard Model Effective Field
  Theory from On-shell Amplitudes}}}, }\href@noop {} {}\,\Eprint
  {http://arxiv.org/abs/1902.06752}{arXiv:1902.06752 [hep-ph]}\BibitemShut
  {NoStop}%
\bibitem [{\citenamefont{Christensen} and
  \citenamefont{Field}(2018)}]{Christensen:2018zcq}%
  \BibitemOpen
  \bibfield{author}{\bibinfo{author}{\bibfnamefont{N.}\,\bibnamefont{Christensen}}
  and \bibinfo{author}{\bibfnamefont{B.}\,\bibnamefont{Field}},
  }\bibfield{title}{\emph {\bibinfo{title}{{Constructive standard model}}},
  }\href {\doibase 10.1103/PhysRevD.98.016014}
  {\bibfield{journal}{\bibinfo{journal}{Phys.
  Rev.}\,}\textbf{\bibinfo{volume}{D98}}\,(\bibinfo{year}{2018})\,\bibinfo{pages}{016014}},
  \Eprint {http://arxiv.org/abs/1802.00448}{arXiv:1802.00448
  [hep-ph]}\BibitemShut {NoStop}%
\bibitem [{\citenamefont{Christensen} \emph
  {et\,al.}(2020)\citenamefont{Christensen}, \citenamefont{Field},
  \citenamefont{Moore}, and \citenamefont{Pinto}}]{Christensen:2019mch}%
  \BibitemOpen
  \bibfield{author}{\bibinfo{author}{\bibfnamefont{N.}\,\bibnamefont{Christensen}},
  \bibinfo{author}{\bibfnamefont{B.}\,\bibnamefont{Field}},
  \bibinfo{author}{\bibfnamefont{A.}\,\bibnamefont{Moore}},  and
  \bibinfo{author}{\bibfnamefont{S.}\,\bibnamefont{Pinto}},
  }\bibfield{title}{\emph {\bibinfo{title}{{2-, 3- and 4-Body Decays in the
  Constructive Standard Model}}}, }\href {\doibase 10.1103/PhysRevD.101.065019}
  {\bibfield{journal}{\bibinfo{journal}{Phys.
  Rev.}\,}\textbf{\bibinfo{volume}{D101}}\,(\bibinfo{year}{2020})\,\bibinfo{pages}{065019}},
  \Eprint {http://arxiv.org/abs/1909.09164}{arXiv:1909.09164
  [hep-ph]}\BibitemShut {NoStop}%
\bibitem [{\citenamefont{Bachu} and
  \citenamefont{Yelleshpur}()}]{Bachu:2019ppp}%
  \BibitemOpen
  \bibfield{author}{\bibinfo{author}{\bibfnamefont{B.}\,\bibnamefont{Bachu}}
  and \bibinfo{author}{\bibfnamefont{A.}\,\bibnamefont{Yelleshpur}},
  }\bibfield{title}{\emph {\bibinfo{title}{{On-Shell Electroweak Sector and the
  Higgs Mechanism}}}, }\href@noop {} {}\,\Eprint
  {http://arxiv.org/abs/1912.04334}{arXiv:1912.04334 [hep-th]}\BibitemShut
  {NoStop}%
\bibitem [{\citenamefont{Aoude} and
  \citenamefont{Machado}(2019)}]{Aoude:2019tzn}%
  \BibitemOpen
  \bibfield{author}{\bibinfo{author}{\bibfnamefont{R.}\,\bibnamefont{Aoude}}
  and \bibinfo{author}{\bibfnamefont{C.~S.} \bibnamefont{Machado}},
  }\bibfield{title}{\emph {\bibinfo{title}{{The Rise of SMEFT On-shell
  Amplitudes}}}, }\href {\doibase 10.1007/JHEP12(2019)058}
  {\bibfield{journal}{\bibinfo{journal}{JHEP}\,}\textbf{\bibinfo{volume}{12}}\,(\bibinfo{year}{2019})\,\bibinfo{pages}{058}},
  \Eprint {http://arxiv.org/abs/1905.11433}{arXiv:1905.11433
  [hep-ph]}\BibitemShut {NoStop}%
\bibitem [{\citenamefont{Durieux} \emph {et\,al.}(2020)\citenamefont{Durieux},
  \citenamefont{Kitahara}, \citenamefont{Shadmi}, and
  \citenamefont{Weiss}}]{Durieux:2019eor}%
  \BibitemOpen
  \bibfield{author}{\bibinfo{author}{\bibfnamefont{G.}\,\bibnamefont{Durieux}},
  \bibinfo{author}{\bibfnamefont{T.}\,\bibnamefont{Kitahara}},
  \bibinfo{author}{\bibfnamefont{Y.}\,\bibnamefont{Shadmi}},  and
  \bibinfo{author}{\bibfnamefont{Y.}\,\bibnamefont{Weiss}},
  }\bibfield{title}{\emph {\bibinfo{title}{{The electroweak effective field
  theory from on-shell amplitudes}}}, }\href {\doibase 10.1007/JHEP01(2020)119}
  {\bibfield{journal}{\bibinfo{journal}{JHEP}\,}\textbf{\bibinfo{volume}{01}}\,(\bibinfo{year}{2020})\,\bibinfo{pages}{119}},
  \Eprint {http://arxiv.org/abs/1909.10551}{arXiv:1909.10551
  [hep-ph]}\BibitemShut {NoStop}%
\bibitem [{\citenamefont{Ruhdorfer} \emph {et\,al.}()\citenamefont{Ruhdorfer},
  \citenamefont{Serra}, and \citenamefont{Weiler}}]{Ruhdorfer:2019qmk}%
  \BibitemOpen
  \bibfield{author}{\bibinfo{author}{\bibfnamefont{M.}\,\bibnamefont{Ruhdorfer}},
  \bibinfo{author}{\bibfnamefont{J.}\,\bibnamefont{Serra}},  and
  \bibinfo{author}{\bibfnamefont{A.}\,\bibnamefont{Weiler}},
  }\bibfield{title}{\emph {\bibinfo{title}{{Effective Field Theory of Gravity
  to All Orders}}}, }\href@noop {} {}\,\Eprint
  {http://arxiv.org/abs/1908.08050}{arXiv:1908.08050 [hep-ph]}\BibitemShut
  {NoStop}%
\bibitem [{\citenamefont{Mangano} and
  \citenamefont{Parke}(1991)}]{Mangano:1990by}%
  \BibitemOpen
  \bibfield{author}{\bibinfo{author}{\bibfnamefont{M.~L.}
  \bibnamefont{Mangano}} and \bibinfo{author}{\bibfnamefont{S.~J.}
  \bibnamefont{Parke}}, }\bibfield{title}{\emph {\bibinfo{title}{{Multiparton
  amplitudes in gauge theories}}}, }\href {\doibase
  10.1016/0370-1573(91)90091-Y} {\bibfield{journal}{\bibinfo{journal}{Phys.
  Rept.}\,}\textbf{\bibinfo{volume}{200}}\,(\bibinfo{year}{1991})\,\bibinfo{pages}{301}},
  \Eprint
  {http://arxiv.org/abs/hep-th/0509223}{arXiv:hep-th/0509223}\BibitemShut
  {NoStop}%
\bibitem [{\citenamefont{Elvang} and
  \citenamefont{Huang}(2015)}]{Elvang:2015rqa}%
  \BibitemOpen
  \bibfield{author}{\bibinfo{author}{\bibfnamefont{H.}\,\bibnamefont{Elvang}}
  and \bibinfo{author}{\bibfnamefont{Y.-t.} \bibnamefont{Huang}}, }\href
  {http://www.cambridge.org/mw/academic/subjects/physics/theoretical-physics-and-mathematical-physics/scattering-amplitudes-gauge-theory-and-gravity?format=AR}
  {\emph {\bibinfo{title}{{Scattering Amplitudes in Gauge Theory and
  Gravity}}}}\,(\bibinfo{publisher}{Cambridge University Press},
  \bibinfo{year}{2015})\BibitemShut {NoStop}%
\bibitem [{\citenamefont{Schwartz}(2014)}]{Schwartz:2013pla}%
  \BibitemOpen
  \bibfield{author}{\bibinfo{author}{\bibfnamefont{M.~D.}
  \bibnamefont{Schwartz}}, }\href
  {http://www.cambridge.org/us/academic/subjects/physics/theoretical-physics-and-mathematical-physics/quantum-field-theory-and-standard-model}
  {\emph {\bibinfo{title}{{Quantum Field Theory and the Standard
  Model}}}}\,(\bibinfo{publisher}{Cambridge University Press},
  \bibinfo{year}{2014})\BibitemShut {NoStop}%
\bibitem [{\citenamefont{McGady} and
  \citenamefont{Rodina}(2014)}]{McGady:2013sga}%
  \BibitemOpen
  \bibfield{author}{\bibinfo{author}{\bibfnamefont{D.~A.} \bibnamefont{McGady}}
  and \bibinfo{author}{\bibfnamefont{L.}\,\bibnamefont{Rodina}},
  }\bibfield{title}{\emph {\bibinfo{title}{{Higher-spin massless $S$-matrices
  in four-dimensions}}}, }\href {\doibase 10.1103/PhysRevD.90.084048}
  {\bibfield{journal}{\bibinfo{journal}{Phys.
  Rev.}\,}\textbf{\bibinfo{volume}{D90}}\,(\bibinfo{year}{2014})\,\bibinfo{pages}{084048}},
  \Eprint {http://arxiv.org/abs/1311.2938}{arXiv:1311.2938
  [hep-th]}\BibitemShut {NoStop}%
\bibitem [{\citenamefont{Durh}(2011)}]{Duhr:2011}%
  \BibitemOpen
  \bibfield{author}{\bibinfo{author}{\bibfnamefont{C.}\,\bibnamefont{Durh}},
  }\bibfield{title}{\emph {\bibinfo{title}{Momentum twistors, special functions
  and symbols}}, }\href
  {https://indico.cern.ch/event/137430/contributions/146026/attachments/113502/161251/Atrani_1_Duhr.pdf}
  {\bibfield{journal}{\bibinfo{journal}{School of analytic computing,
  Atrani}}\,(\bibinfo{year}{7 Oct 2011})}\BibitemShut {NoStop}%
\bibitem [{\citenamefont{Henning} \emph
  {et\,al.}(2017{\natexlab{b}})\citenamefont{Henning}, \citenamefont{Lu},
  \citenamefont{Melia}, and \citenamefont{Murayama}}]{Henning:2017fpj}%
  \BibitemOpen
  \bibfield{author}{\bibinfo{author}{\bibfnamefont{B.}\,\bibnamefont{Henning}},
  \bibinfo{author}{\bibfnamefont{X.}\,\bibnamefont{Lu}},
  \bibinfo{author}{\bibfnamefont{T.}\,\bibnamefont{Melia}},  and
  \bibinfo{author}{\bibfnamefont{H.}\,\bibnamefont{Murayama}},
  }\bibfield{title}{\emph {\bibinfo{title}{{Operator bases, $S$-matrices, and
  their partition functions}}}, }\href {\doibase 10.1007/JHEP10(2017)199}
  {\bibfield{journal}{\bibinfo{journal}{JHEP}\,}\textbf{\bibinfo{volume}{10}}\,(\bibinfo{year}{2017}{\natexlab{b}})\,\bibinfo{pages}{199}},
  \Eprint {http://arxiv.org/abs/1706.08520}{arXiv:1706.08520
  [hep-th]}\BibitemShut {NoStop}%
\bibitem [{\citenamefont{Li} \emph {et\,al.}()\citenamefont{Li},
  \citenamefont{Ren}, \citenamefont{Shu}, \citenamefont{Xiao},
  \citenamefont{Yu}, and \citenamefont{Zheng}}]{Li:2020gnx}%
  \BibitemOpen
  \bibfield{author}{\bibinfo{author}{\bibfnamefont{H.-L.} \bibnamefont{Li}},
  \bibinfo{author}{\bibfnamefont{Z.}\,\bibnamefont{Ren}},
  \bibinfo{author}{\bibfnamefont{J.}\,\bibnamefont{Shu}},
  \bibinfo{author}{\bibfnamefont{M.-L.} \bibnamefont{Xiao}},
  \bibinfo{author}{\bibfnamefont{J.-H.} \bibnamefont{Yu}},  and
  \bibinfo{author}{\bibfnamefont{Y.-H.} \bibnamefont{Zheng}},
  }\bibfield{title}{\emph {\bibinfo{title}{{Complete Set of Dimension-8
  Operators in the Standard Model Effective Field Theory}}}, }\href@noop {}
  {}\,\Eprint {http://arxiv.org/abs/2005.00008}{arXiv:2005.00008
  [hep-ph]}\BibitemShut {NoStop}%
\bibitem [{\citenamefont{Gripaios} and
  \citenamefont{Sutherland}(2019)}]{Gripaios:2018zrz}%
  \BibitemOpen
  \bibfield{author}{\bibinfo{author}{\bibfnamefont{B.}\,\bibnamefont{Gripaios}}
  and \bibinfo{author}{\bibfnamefont{D.}\,\bibnamefont{Sutherland}},
  }\bibfield{title}{\emph {\bibinfo{title}{{DEFT: A program for operators in
  EFT}}}, }\href {\doibase 10.1007/JHEP01(2019)128}
  {\bibfield{journal}{\bibinfo{journal}{JHEP}\,}\textbf{\bibinfo{volume}{01}}\,(\bibinfo{year}{2019})\,\bibinfo{pages}{128}},
  \Eprint {http://arxiv.org/abs/1807.07546}{arXiv:1807.07546
  [hep-ph]}\BibitemShut {NoStop}%
\bibitem [{\citenamefont{Criado}(2019)}]{Criado:2019ugp}%
  \BibitemOpen
  \bibfield{author}{\bibinfo{author}{\bibfnamefont{J.~C.}
  \bibnamefont{Criado}}, }\bibfield{title}{\emph {\bibinfo{title}{{BasisGen:
  automatic generation of operator bases}}}, }\href {\doibase
  10.1140/epjc/s10052-019-6769-5} {\bibfield{journal}{\bibinfo{journal}{Eur.
  Phys.
  J.}\,}\textbf{\bibinfo{volume}{C79}}\,(\bibinfo{year}{2019})\,\bibinfo{pages}{256}},
  \Eprint {http://arxiv.org/abs/1901.03501}{arXiv:1901.03501
  [hep-ph]}\BibitemShut {NoStop}%
\bibitem [{\citenamefont{Hays} \emph {et\,al.}(2019)\citenamefont{Hays},
  \citenamefont{Martin}, \citenamefont{Sanz}, and
  \citenamefont{Setford}}]{Hays:2018zze}%
  \BibitemOpen
  \bibfield{author}{\bibinfo{author}{\bibfnamefont{C.}\,\bibnamefont{Hays}},
  \bibinfo{author}{\bibfnamefont{A.}\,\bibnamefont{Martin}},
  \bibinfo{author}{\bibfnamefont{V.}\,\bibnamefont{Sanz}},  and
  \bibinfo{author}{\bibfnamefont{J.}\,\bibnamefont{Setford}},
  }\bibfield{title}{\emph {\bibinfo{title}{{On the impact of dimension-eight
  SMEFT operators on Higgs measurements}}}, }\href {\doibase
  10.1007/JHEP02(2019)123}
  {\bibfield{journal}{\bibinfo{journal}{JHEP}\,}\textbf{\bibinfo{volume}{02}}\,(\bibinfo{year}{2019})\,\bibinfo{pages}{123}},
  \Eprint {http://arxiv.org/abs/1808.00442}{arXiv:1808.00442
  [hep-ph]}\BibitemShut {NoStop}%
\bibitem [{\citenamefont{Fonseca}(2020)}]{Fonseca:2019yya}%
  \BibitemOpen
  \bibfield{author}{\bibinfo{author}{\bibfnamefont{R.~M.}
  \bibnamefont{Fonseca}}, }\bibfield{title}{\emph {\bibinfo{title}{{Enumerating
  the operators of an effective field theory}}}, }\href {\doibase
  10.1103/PhysRevD.101.035040} {\bibfield{journal}{\bibinfo{journal}{Phys.
  Rev.}\,}\textbf{\bibinfo{volume}{D101}}\,(\bibinfo{year}{2020})\,\bibinfo{pages}{035040}},
  \Eprint {http://arxiv.org/abs/1907.12584}{arXiv:1907.12584
  [hep-ph]}\BibitemShut {NoStop}%
\bibitem [{\citenamefont{Falkowski}()}]{Falkowski:2019zdo}%
  \BibitemOpen
  \bibfield{author}{\bibinfo{author}{\bibfnamefont{A.}\,\bibnamefont{Falkowski}},
  }\bibfield{title}{\emph {\bibinfo{title}{{Bases of massless EFTs via momentum
  twistors}}}, }\href@noop {} {}\,\Eprint
  {http://arxiv.org/abs/1912.07865}{arXiv:1912.07865 [hep-ph]}\BibitemShut
  {NoStop}%
\end{thebibliography}%

\end{document}